\documentclass[traditabstract,printer]{aa}
\usepackage{natbib}
\usepackage{txfonts}
\usepackage{graphicx}
\bibpunct{(}{)}{;}{a}{}{,}
\begin{document}
\title{Building the Galactic halo from globular clusters: evidence from chemically unusual red giants}
 
\bigskip
\author{S. L. Martell \inst{1} 
\and J. P. Smolinski \inst{2,3}
\and T. C. Beers \inst{2}
\and E. K. Grebel \inst{1}} 
\institute{Astronomisches Rechen-Institut\\
Zentrum f\"{u}r Astronomie der Universit\"{a}t Heidelberg\\
69120 Heidelberg, Germany\\
\email{martell@ari.uni-heidelberg.de, grebel@ari.uni-heidelberg.de}
\and Department of Physics and Astronomy and JINA (Joint Institute for Nuclear Astrophysics)\\
Michigan State University\\
East Lansing, MI 48824, USA\\
\email{smolin19@msu.edu, beers@pa.msu.edu}
\and Department of Physics and Astronomy\\
State University of New York College at Oneonta\\
Oneonta, NY 13820, USA}

\date{Received 7 July 2011 / Accepted 31 August 2011}

\abstract{We present a spectroscopic search for halo field stars that
originally formed in globular clusters. Using moderate-resolution
SDSS-III/SEGUE-2 spectra of $561$ red giants with typical halo
metallicities ($-1.8 \le $[Fe/H] $\le -1.0$), we identify $16$ stars,
$3\%$ of the sample, with CN and CH bandstrength behavior indicating
depleted carbon and enhanced nitrogen abundances relative to the rest of
the data set. Since globular clusters are the only environment known in
which stars form with this pattern of atypical light-element
abundances, we claim that these stars are second-generation globular
cluster stars that have been lost to the halo field via normal cluster
mass-loss processes. Extrapolating from theoretical models of
two-generation globular cluster formation, this result suggests that
globular clusters contributed significant numbers of stars to the
construction of the Galactic halo: we calculate that a minimum of
$17$\% of the present-day mass of the stellar halo was originally
formed in globular clusters. The ratio of CN-strong to CN-normal stars
drops with Galactocentric distance, suggesting that the inner-halo
population may be the primary repository of these stars.}


\titlerunning{Globular cluster stars in the Halo}
\maketitle

\section{Introduction}
Light-element abundance inhomogeneities in globular clusters have been a
subject of ongoing study for over thirty years. Early observations of a
bimodal distribution of CN bandstrength in globular cluster red-giant
(RGB) stars, anticorrelated with CH bandstrength (e.g., Norris
\& Cottrell 1979\nocite{NC79}; Suntzeff 1981\nocite{S81}; Norris et
al. 1984\nocite{N84}), were interpreted as evidence of anticorrelated
ranges of carbon and nitrogen abundances in globular cluster stars
(e.g., Bell \& Dickens 1980\nocite{BD80}). Since that time,
observations have been extended to include more clusters, fainter
stars within individual clusters, and additional elemental abundances. We
now know that anticorrelated C-N, O-Na and Mg-Al ranges are present in
stars on the red-giant branches of all globular clusters that have been
thoroughly studied (e.g., Kayser et al. 2008\nocite{KHG08}; Carretta et
al. 2009\nocite{CBG09}; Smolinski et al. 2011b\nocite{SM11}), and
that the C-N anticorrelation is also found among main-sequence stars in
several clusters (e.g., Harbeck et al. 2003\nocite{HSG03}; Briley et al.
2004\nocite{BCS04}; Pancino et al. 2010\nocite{PR10}; Smolinski et
al. 2011b\nocite{SM11}). Light-element abundances in globular cluster
stars range from scaled-Solar, similar to halo field stars of the same
metallicity, to depleted in C, O and Mg and enhanced in N, Na and Al,
with between 30\% \citep{PR10} and 70\% \citep{CBG09} of stars in a
given cluster having an atypical abundance pattern.

The atypical abundance pattern resembles the result of
high-temperature hydrogen burning: the CNO cycle converts
carbon and oxygen into nitrogen, and the NeNa and MgAl cycles increase
the abundances of sodium and aluminium while depleting magnesium. The
presence of this abundance pattern in old, low-mass main-sequence stars, which
have neither the high-temperature fusion zones nor the capacity to
transport material between their cores and surfaces, requires that the
abundance variations be extrinsic. The fact that the abundance
differences between stars with typical and atypical abundance patterns
are equally large before and after first dredge-up, which occurs low on the RGB,
indicates that the abundance differences are present throughout the
stars, and are not merely surface pollution.

As a result, the prevailing model for the origin of primordial
light-element abundance variations in globular clusters is that the
stars with atypical light-element abundances are a nearly coeval
second generation formed from material processed by intermediate- or
high-mass stars in the first generation. There are several possible
sites for high-temperature fusion processing in the first
generation. AGB stars are a common suggestion (e.g., Parmentier et
al. 1999\nocite{PJM99}) because they have relatively slow winds, they
are a site of hot hydrogen burning (e.g., Karakas 2010\nocite{K10}), and
they evolve on timescales of $\sim 10^{8}$ years, quite fast compared to
the lifetime of a globular cluster. Other possible sources for
high-temperature-processed feedback material are rapidly rotating
massive stars \citep{DM07} and massive binary stars undergoing mass
transfer \citep{DMP09}, both of which could return more feedback mass in
a shorter amount of time than AGB stars, although the stars
themselves are less common. 

\begin{figure}
\resizebox{\hsize}{!}{\includegraphics{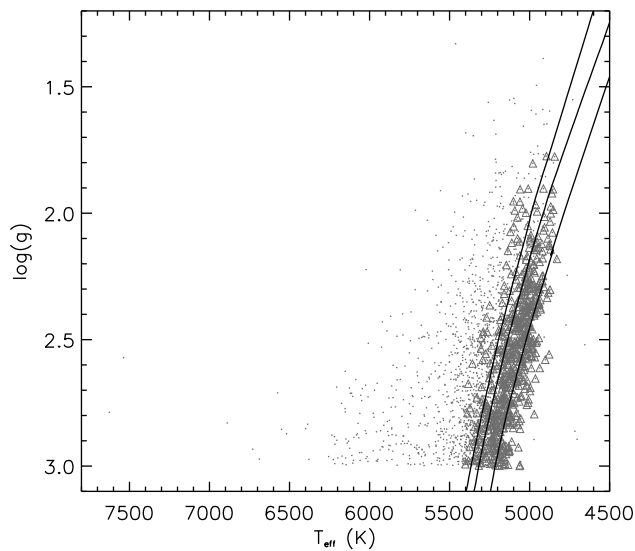}}
\caption[]{Surface gravity versus effective temperature for our
  initial (small gray points) and trimmed (open gray
  triangles) data sets. Solid black lines are 12 Gyr Dartmouth
  isochrones with (left to right) [Fe/H]$=-1.8$, $-1.4$, and $-1.0$.}
\end{figure}

However, all of these possible feedback sources suffer from what has
come to be known as the ``mass budget problem,'' the mismatch between
the current 1:1 ratio of first- to second-generation stars and the
concept that the second generation was built entirely from winds from
the first generation. Recent theoretical models for globular
cluster formation propose two solutions to the mass budget
problem: massive first generations and significant gas accretion. The
models of \citet{BC07} and \citet{DVD08} both require a first
generation with an initial mass $10$ to $20$ times its present mass,
and have second generations built entirely from AGB winds. In the
model of \citet{DVD08}, the second-generation star formation occurs
near the center of the cluster, and concludes when type Ia supernovae
begin. The Ia SNe drive an overall expansion of the cluster, causing
stars at large radii (primarily first-generation stars) to dissociate
from the cluster, thereby reducing the ratio of first- to second-generation
stars to present-day levels. AGB stars are also the source of chemical
differences between the first and second generation in the model of
\citet{CS11}, but their model involves a significant amount of gas
being accreted from the environment, reducing the mass required
for the first generation of stars. The authors investigate the
mass-accretion rate for proto-globular clusters in a variety of
cosmological environments, and conclude that clusters with initial
masses above $10^{4}{\rm M_{\odot}}$ would have been able to accrete
significant amounts of gas in the early Milky Way. They also calculate
that the minimum mass for sufficient gas accretion would be lower for
globular clusters forming as satellites of dwarf galaxies, a claim that could
be tested with observations of intermediate-age star clusters in the
Large Magellanic Cloud (LMC). Interestingly, the recent
high-resolution study of \citet{MCB11} finds no sign of light-element
anticorrelations among 14 red giant stars in the massive
intermediate-age LMC cluster NGC 1866. The authors suggest that the
minimum mass for self-enrichment in the LMC is on the order of
$10^{5}{\rm M_{\odot}}$. Although there are questions about the
likelihood that early globular clusters would accrete unprocessed gas
from their surroundings with precisely the right metallicity (e.g.,
Martell 2011\nocite{M11}), and the gas dynamics involved in mass loss
and dilution in globular clusters have been the subject of few
advanced numerical studies (e.g., Priestley et al. 2011\nocite{PR11};
D'Ercole et al. 2011\nocite{DDV11}), the \citet{CS11} model provides
an important discussion of the interplay between early cluster
self-enrichment and the galaxy-scale environment.

\begin{figure}
\resizebox{\hsize}{!}{\includegraphics{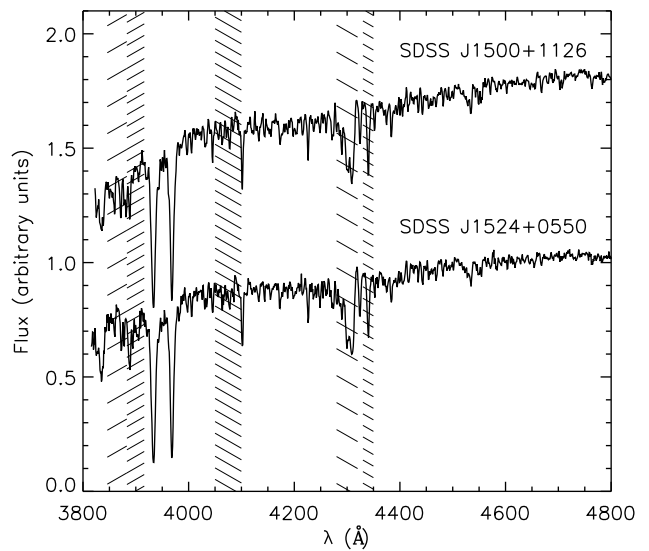}}
\caption[]{Two sample spectra from our data set, demonstrating the
  effects of varying [C/Fe] and [N/Fe] at fixed stellar
  parameters. The upper spectrum, of SDSS J1500+1126, is CN-strong
  while the lower spectrum, of SDSS J1524+0550, is CN-normal. Shaded
  regions denote the bandpasses of $S(3839)$ and $S(CH)$: lines angled
  up to the right mark the sideband (more closely spaced lines) and
  science band (more broadly spaced lines) of $S(3839)$, and lines
  angled down to the right mark the sidebands (more closely spaced
  lines) and science band (more broadly spaced lines) of $S(CH)$.}
\end{figure}

These scenarios rely implicitly on the high mass and high density of
early globular clusters to retain and/or accrete gas well enough to
permit a second burst of star formation. It is therefore not surprising
that although stars with second-generation-like light-element abundances
are found in every globular cluster in the Milky Way, they are not found
at all in old open clusters. In studies of individual stars (e.g.,
Jacobson et al. 2008\nocite{JFP08}; Martell \& Smith 2009\nocite{MS09})
and mean abundances (e.g., de Silva et al. 2009\nocite{dSG09}) for old
open clusters, light-element abundance behavior is found to be
distinctly different from what is observed in globular clusters. They
are also quite {\it unlikely} to have formed in the halo field,
but they have recently been found there: \citet{MG10} (hereafter MG10)
searched the SEGUE survey (Sloan Extension for Galactic Understanding
and Exploration, Yanny et al. 2009\nocite{YR09}) for halo giants with
unusually strong UV/blue CN bands and identified $49$ (of roughly
$2000$) stars likely to have low carbon abundances and high nitrogen
abundances. Similarly, \citet{CBG10b} compiled a sample of 144 metal-poor disk,
halo and bulge stars from the literature and identified 2 of those as
Na-rich and likely to have originated in globular clusters. In both
cases, the authors interpret the field stars with second-generation
abundances as stars that formed in globular clusters and were later
transferred to the halo through mass-loss processes such as tidal
stripping by the Galaxy or two-body interactions within the cluster. It
is also claimed in MG10 that their figure of $2.5\%$ of halo stars
having second-generation abundances, seen through the lens of a globular
cluster formation model with strong early mass loss (such as, e.g.,
D'Ercole et al. 2008\nocite{DVD08}), implies that as much as $50\%$ of
halo field stars originally formed in globular clusters. \citet{CBG10b}
conclude that early mass loss from globular clusters is ``a major
building block of the halo'', and may also have contributed
significantly to the formation of the thick disk. This possibility has
been suggested theoretically as well: both \citet{BK08} and \citet{MK10}
find that early mass loss from globular clusters should contribute
significant numbers of stars to the halo field.

\begin{figure}
\resizebox{\hsize}{!}{\includegraphics{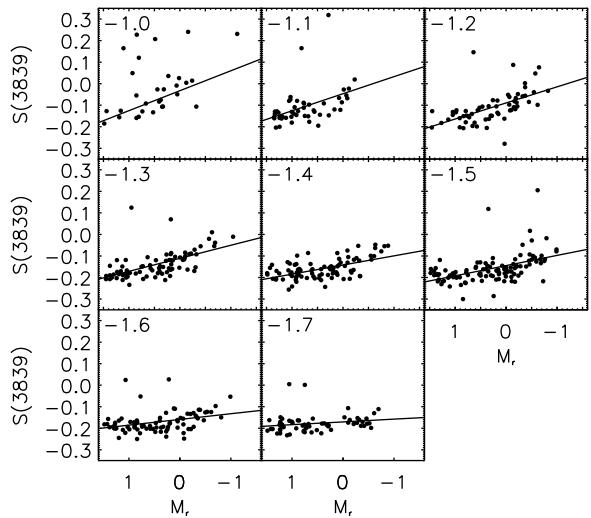}}
\caption[]{CN bandstrength index $S(3839)$ versus absolute $r$
  magnitude for our full data set, divided into $0.1$-dex-wide bins in
[Fe/H]. Maximum [Fe/H] in each bin is given in the upper left corner
of each panel. Solid lines are best fits to the CN-normal stars in
each panel. The slope of the best-fit line and the separation between
CN-normal and CN-strong stars both decline with decreasing metallicity.}
\end{figure}

In this paper we use data from the SEGUE-2 moderate-resolution
spectroscopic survey, available as part of the eighth data release of
the Sloan Digital Sky Survey (Aihara et al. 2011\nocite{DR8};
Eisenstein et al. 2011\nocite{SDSSIII}; York et al. 2000\nocite{Y00}),
to search for CN-strong halo giants like those identified in
MG10. This will allow us to make two important extensions to that
work: first, by expanding the sample of stars surveyed, we can refine
our estimate of ${\rm f_{h}^{2G}}$, the present-day fraction of
CN-strong stars in the halo field; second, we calculate ${\rm f_
    {h}^{GC}}$, the fraction of halo stars with first- or
second-generation chemistry originating in globular clusters, as
predicted by several different two-generation cluster formation
models.

\section{The Data Set}

The SDSS and its extensions have acquired $ugriz$ photometry for
several hundred million stars. SEGUE, one of three sub-surveys that
together formed SDSS-II, extended the $ugriz$ imaging footprint of
SDSS-I \citep{FI96,GC98,Y00,PM03,GS06, SL02,DR1,DR2,DR3,DR4,DR5,DR6,DR7}
by approximately 3500 deg${\rm ^{2}}$, and also obtained $R \simeq 2000$
spectroscopy for approximately $240,000$ stars over a wavelength range
of $3800-9200\hbox{\AA}$. SEGUE included spectra for a collection of
Galactic globular and open clusters, which served as calibrators for the
${\rm T_{eff}}$, ${\rm log(g)}$, and [Fe/H] scales for all stars
observed by SDSS/SEGUE, as processed by the SEGUE Stellar Parameter
Pipeline (SSPP; Lee et al. 2008a; b\nocite{LBS08a}\nocite{LBS08b};
Allende Prieto et al. 2008\nocite{AP08}). The SSPP produces estimates of
${\rm T_{eff}}$, ${\rm log(g)}$, [Fe/H], and radial velocity, along with
the equivalent widths and/or line indices for $85$ atomic and molecular
absorption lines, by processing the calibrated spectra generated by the
standard SDSS spectroscopic reduction pipeline \citep{SL02}. See
\citet{LBS08a} for a detailed discussion of the approaches used by the
SSPP; \citet{SL11} provides details on the most recent updates to this
pipeline, along with additional validations. SEGUE-2 (C. Rockosi et al.,
in prep.) expanded the numbers of Galactic stars with available
low-resolution SDSS spectra by over $120,000$, concentrating in
particular on stars with greater distances in order to better sample the
outer-halo region of the Galaxy.

\begin{figure}
\resizebox{\hsize}{!}{\includegraphics{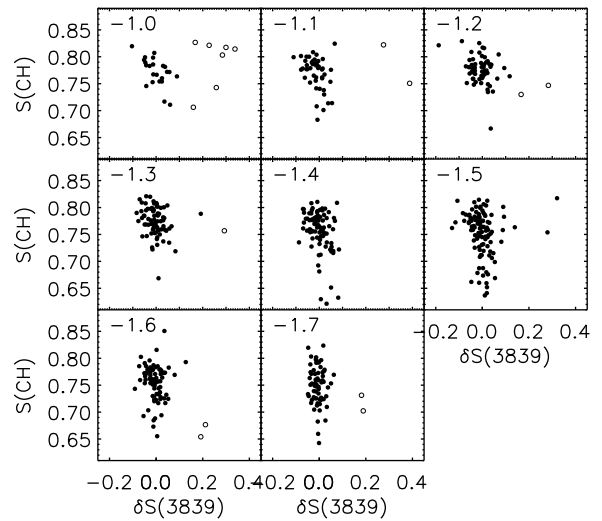}}
\caption[]{CN-CH planes for our final data set, divided into $0.1$-dex-wide bins in [Fe/H]. The CH bandstrength index $S(CH)$ is plotted versus the CN bandstrength residual $\delta S(3839)$, and candidate globular cluster stars are drawn as open circles. As in Figure 3, the maximum metallicity in each bin is given in the upper left corner. As is explained in the text, we select as candidate globular cluster stars all those with relatively high $\delta S(3839)$ in the three highest-metallicity bins, and stars with both high $\delta S(3839)$ and $S(CH)$ below the mean $S(CH)$ of the CN-normal group in all other metallicity bins.}
\end{figure}

As in MG10, SEGUE-2 targets were selected from across a large range in
parameter space: we began by selecting all SDSS-III targets from
SEGUE-2 plates with [Fe/H]$\leq -1.0$, ${\rm log(g)}\leq 3.0$,
$(g-r)_{0}\geq 0.2$, and a mean signal-to-noise ratio (SNR) of at
least $20$ per pixel. Errors on the derived quantities and
quality-assurance flags were also required to fall within a certain
range: $\sigma_{\rm [Fe/H]}\leq 0.5$, $\sigma_{\rm log(g)}\leq 0.5$,
$\sigma_{\rm Teff}\leq 200$~K. This initial data set contained 2019
stars. We then identified four stars as carbon-rich using the C$_{2}$
indices defined in MG10, and removed them from the sample. RGB
stars were identified by dividing the data into $0.2$-dex-wide
metallicity bins and making a recursive 3-sigma selection about a
fiducial sequence in the $({\rm log(g)}$, ${\rm T_{eff}})$ plane. Since
our analysis is based on the CN and CH molecular absorption features,
which become quite weak at low metallicity, we also restrict our sample
to stars with [Fe/H] $\geq -1.8$ and a SNR per
pixel between $4000\hbox{\AA}$ and $4100\hbox{\AA}$ (SN$_{\rm blue}$)
above $15$. The RGB, metallicity, and SN$_{\rm blue}$ selections together
reduce our data set to $561$ stars. 

\begin{table*}
\caption{Star identifiers, position, photometry and parameters for all
  program stars from this study and from MG10.}
\centering
\begin{tabular}{c c c c c c c c c c c c}
\hline \hline
SDSS ID & Plate & MJD & FiberID & $\alpha$ & $\delta$ & $g_{0}$ &
$(g-r)_{0}$ & [Fe/H] & ${\rm T}_{eff}$ & log(g) & Survey\\ 
\hline
SDSS J1605+1638 & 3289 & 54910 & 132 & 16:05:52.72 & 16:38:59.21 & 16.04 & 0.663 & -1.07 & 4965 & 2.11 & SEGUE-2\\
SDSS J0842+2216 & 3373 & 54940 & 51 & 08:42:16.23 & 22:16:58.27 & 16.14 & 0.715 & -1.05 & 4913 & 2.06 & SEGUE-2\\
SDSS J1528+0759 & 3308 & 54919 & 529 & 15:28:51.34 & 07:59:08.62 & 17.11 & 0.696 & -1.04 & 4950 & 2.29 & SEGUE-2\\
SDSS J1557+3629 & 3478 & 55008 & 229 & 15:57:48.95 & 36:29:33.04 & 16.77 & 0.735 & -1.03 & 4910 & 2.29 & SEGUE-2\\
SDSS J1011+1810 & 3178 & 54848 & 269 & 10:11:13.50 & 18:10:32.73 & 17.22 & 0.689 & -1.00 & 4934 & 2.37 & SEGUE-2\\
SDSS J2243+1322 & 3128 & 54776 & 104 & 22:43:31.49 & 13:22:33.68 & 17.72 & 0.689 & -1.09 & 4877 & 2.55 & SEGUE-2\\
SDSS J1016+1737 & 3178 & 54848 & 55 & 10:16:42.76 & 17:37:23.51 & 17.44 & 0.619 & -1.05 & 5066 & 2.50 & SEGUE-2\\
SDSS J1206+3021 & 3181 & 54860 & 73 & 12:06:49.60 & 30:21:49.39 & 18.06 & 0.739 & -1.05 & 4983 & 2.53 & SEGUE-2\\
SDSS J1210+1042 & 3214 & 54866 & 14 & 12:10:11.17 & 10:42:28.64 & 16.80 & 0.652 & -1.07 & 5025 & 2.56 & SEGUE-2\\
SDSS J1617+0502 & 3298 & 54924 & 48 & 16:17:42.83 & 05:02:32.70 & 17.54 & 0.663 & -1.06 & 4991 & 2.56 & SEGUE-2\\
\hline
\end{tabular}
\end{table*}

Some stars from SEGUE-1 were intentionally re-observed in SEGUE-2,
as a way to check for consistency in data reduction and analysis between
the two surveys. However, none of the stars in this study were also in
the MG10 data set. Table 1 lists SDSS identifiers, the plug plate
number, MJD, and fiber number of the SEGUE observation (useful for
selecting individual stars from the SDSS Catalog Archive
Server\footnote{http://skyservice.pha.jhu.edu/CasJobs/login.aspx}),
right ascension, declination, photometry, SSPP-derived stellar
parameters, and survey name (SEGUE-1 or SEGUE-2) for each star in the
final trimmed data set and the final data set from MG10. Only the
first ten rows of Table 1 are given in the print version of this
paper, to indicate form and content; a full version is available at the
CDS \footnote{via anonymous ftp to cdsarc.u-strasbg.fr (130.79.128.5)
  or via http://cdsweb.u-strasbg.fr/cgi-bin/qcat?J/A+A/}. Figure 1 shows surface gravity
versus effective temperature for our initial (small gray points) and
trimmed (open gray triangles) data sets, using the ${\rm T_{eff}}$ and
${\rm log(g)}$ values derived from the spectra by the SSPP. The
overplotted solid lines are 12 Gyr Dartmouth isochrones \citep{DC08}
with metallicities of [Fe/H]$=-1.8$, $-1.4$, and $-1.0$, spanning the
metallicity range of our data set.

\section{Analysis}

Because we are searching for old, low-mass red giant stars with
globular-cluster-like abundance patterns, our analysis follows a
process typical in the globular cluster literature. Specifically, we are
using index-based techniques developed for globular cluster studies as
indicators of carbon and nitrogen abundance, and we infer the
presence of the full second-generation light-element abundance pattern
from those indices. Overall, we followed the procedure from
MG10 fairly closely in analyzing our data, in order for the results to
be as compatible as possible. 

\subsection{Bandstrength indices}

We use the bandstrength indices $S(3839)$ \citep{N81} and $S(CH)$
\citep{MSB08}, which were initially developed for studies of
light-element abundance variations in globular clusters, to identify
stars with strong CN bands and weak CH bands relative to the majority of
the halo field. Indices measure the magnitude difference between the
integrated flux in the feature (the ``science band'') and one or two
nearby, independent regions of the spectrum (the ``sidebands''), in the
sense that more absorption in the feature results in a larger index
value. Since molecular abundance is strongly controlled by the abundance
of the minority species, the CH band is indicative of [C/Fe] while the
CN band traces [N/Fe]. We are using carbon and nitrogen abundances as an
indicator of the ``second-generation'' abundance pattern of depleted
carbon, oxygen, and magnesium and enhanced nitrogen, sodium, and aluminum
because that full abundance pattern is consistently found together in
Galactic globular clusters.

Figure 2 shows typical spectra from our data set, chosen to illustrate
the range of variation in CN and CH bandstrength. The upper spectrum
(of SDSS J1500+1126) is of a CN-strong star with ${\rm T_{eff}}=5250$,
[Fe/H]$=-1.70$, $S(3839)=0.005$ and $S(CH)=0.70$, and the lower
spectrum (of SDSS J1524+0550) has similar stellar parameters but
relatively weaker CN and stronger CH bands (${\rm T_{eff}}=5210$,
[Fe/H]$=-1.74$, $S(3839)=-0.189$ and $S(CH)=0.79$). The spectra are
both normalized near $4500\hbox{\AA}$ and offset vertically. The
shaded areas with lines tipping up to the right mark the sideband
(more closely spaced lines) and science band (more broadly spaced lines) of
$S(3839)$, and the shaded areas with lines angled down to the right
mark the sidebands (more closely spaced lines) and science band (more
broadly spaced lines) of $S(CH)$. The bandpasses of both indices are
given in Table 2. Errors on the bandstrengths were estimated as in
MG10, using a Monte Carlo technique to randomly sample the SEGUE noise
vectors and add that noise to the spectra, then measure the standard
deviation of the bandstrength over 100 realizations. Typical values
for $\sigma_{S(3839)}$ and $\sigma_{S(CH)}$ are $0.015$ and $0.007$,
respectively.

\subsection{Identification of CN-Strong candidates}

We identify candidate second-generation halo stars in our data set in
two stages, first by selecting those stars with unusually high $S(3839)$
relative to other stars of similar metallicity and evolutionary stage,
and then by taking just the subset of those CN-strong stars with
appropriate $S(CH)$. Absolute magnitudes for our stars were calculated
using a spectroscopic parallax method similar to that used in MG10. We
used 12-Gyr Dartmouth isochrones \citep{DC08}, interpolated to the
metallicities of our target stars, then use dereddened $(g-r)_{0}$
colors to predict absolute ${\rm M}_{r}$ magnitudes.

Figure 3 shows $S(3839)$ versus ${\rm M}_{r}$ for our 561 program stars,
divided into $0.1$-dex-wide metallicity bins. The solid lines
overplotted are linear fits to the CN-normal stars, and the maximum
metallicity in each bin is given in the upper right corner of each
panel. The distinctive characteristics of each panel are the dominance
of the CN-normal group, the slope of the best-fit line, and the
separation of the CN-normal and CN-strong groups. In globular
clusters, the fraction of CN-strong stars is roughly $50\%$ \citep{K94}, but the
slope of the CN bandstrength-luminosity relation and the separation in CN
bandstrength between the two groups behaves similarly to what we
observe in field stars. As in MG10, the slope and the separation both
decline with decreasing metallicity, and the two CN bandstrength
groups become difficult to distinguish at the lowest metallicites. 

\begin{figure}
\resizebox{\hsize}{!}{\includegraphics{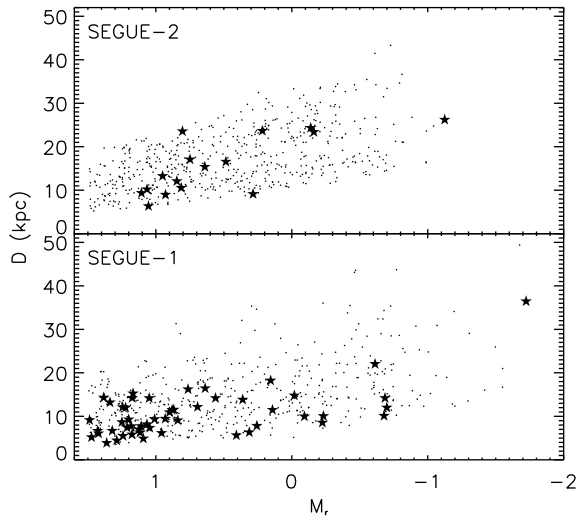}}
\caption[]{Heliocentric distance as a function of absolute $r$ magnitude for our final data set (upper panel) and for the MG10 data set (lower panel). CN-normal stars are plotted as small dots, while CN-strong stars are shown as filled stars. Only a randomly selected $25\%$ of the CN-normal MG10 stars is shown to reduce crowding. In both panels, the lower boundary is set by the bright limit of the SEGUE survey ($g=14$), and the upper boundary is set by our requirement that ${\rm SN_{blue}}>15$.}
\end{figure}

We measure the quantity $\delta S(3839)$, the difference between an
individual star's CN bandstrength and the best-fit line at the same
absolute magnitude, in the plane of Figure 3. The eight panels of
Figure 4 correspond to the panels of Figure 3, and show $\delta
S(3839)$ versus $S(CH)$ for each metallicity bin, with candidate
second-generation field stars plotted as open circles and all other stars
plotted as filled circles. Because the $S(CH)$ index has a reduced
sensitivity to carbon abundance at the upper end of our metallicity
range, we accept all CN-strong stars with metallicities above
[Fe/H]$=-1.3$ as candidates, and we select only those CN-strong stars
with $S(CH)$ below the mean of the CN-normal group for stars with
metallicity below [Fe/H]$=-1.3$. These are the same criteria as were
used in MG10, and they return $16$ candidate globular cluster stars,
$3\%$ of our sample. This is very similar to the MG10 result based on
SEGUE-I stars. 

\subsection{Stellar distances}

We calculate heliocentric distances D to our target stars from the
distance modulus $(r_{0}-{\rm M}_{r})$. Errors in distance were
calculated by randomly sampling the error on $(g-r)_{0}$, then
recalculating the distance. The standard deviation in $100$ of these
realizations was then adopted as $\sigma_{\rm D}$. One of the goals of
the SEGUE-2 survey was to observe stars at greater distances than were
observed in SEGUE-1, and while our data set covers a heliocentric
distance range very similar to that of MG10, the average star at any
fixed ${\rm M}_{r}$ is more distant for the current data set than for
the MG10 data set. The upper panel of Figure 5 shows our calculated
distances versus absolute ${\rm M}_{r}$ magnitudes for the SEGUE-2 data
set, with CN-normal stars drawn as small dots and CN-strong stars drawn
as filled stars. The lower panel shows the analogous data for the MG10
data set, with only a randomly selected $25\%$ of the MG10 CN-normal
stars plotted for visual clarity. In both panels, the stars fill a band
that is restricted at small distances by the SEGUE bright limit of
$g=14$ and at large distances by our requirement that SN$_{\rm blue}
\geq 15.0$. 

\begin{figure}
\resizebox{\hsize}{!}{\includegraphics{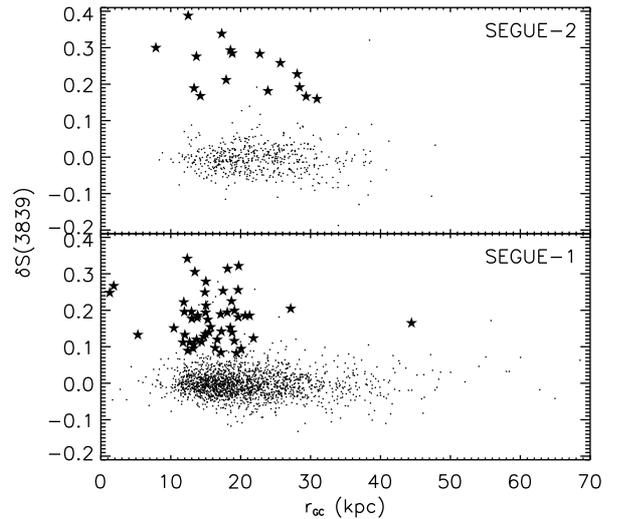}}
\caption[]{CN bandstrength index $\delta S(3839)$ versus
  Galactocentric distance ${\rm r_{GC}}$ for our final data set (upper
  panel) and the MG10 sample (lower panel). Symbols are the same as in
  Figure 5. There is a sharp drop in the number and frequency of
  CN-strong stars at roughly $20$ kpc. }
\end{figure}

Heliocentric distances are converted geometrically to Galactocentric
distances ${\rm r_{GC}}$, using the relation ${\rm
r_{GC}^{2}=D^{2}+(8\hbox{kpc})^{2}-2D\cos(\gamma)}$, in which $\gamma$
is the angle on the sky between the Galactic center and the star in
question. Table 3 lists plate, MJD, and FiberID identifiers, along
with our measurements of $S(3839)$ and $S(CH)$, our determination of a
star's CN class (normal or strong), absolute ${\rm M}_{r}$ magnitude,
heliocentric distance, Galactocentric distance, and survey name (SEGUE-1
or SEGUE-2), for the same stars as in Table 1. As with Table 1, only
the first ten rows are given in the print version of this paper, and a
full version of the table is available from CDS. Figure 6 shows
$\delta S(3839)$ versus ${\rm r_{GC}}$ for our final data set (upper
panel) and the MG10 sample (lower panel), using the same symbols as in
Figure 5. The frequency of CN-strong stars appears to drop at roughly
$20$ kpc, with only one CN-strong star beyond ${\rm r_{GC}}=30$ kpc. 

\begin{figure}
\resizebox{\hsize}{!}{\includegraphics{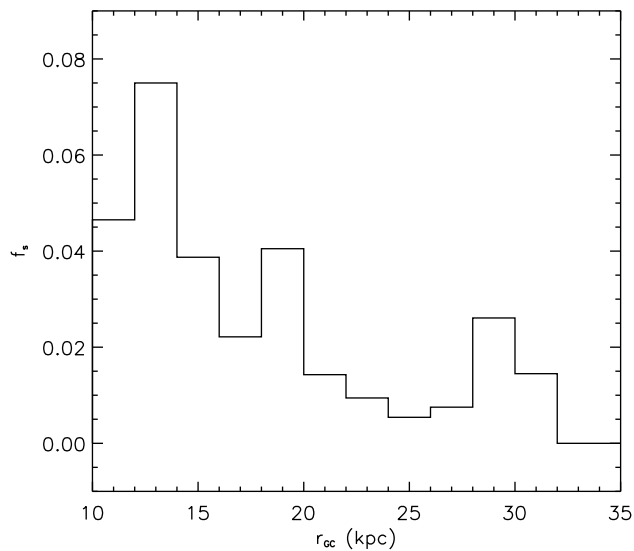}}
\caption[]{The ratio of CN-strong to CN-normal stars, ${\rm f}_{s}$, as a function of Galactocentric distance for the present data set and the MG10 data combined. Vertical error bars represent the Poisson errors on ${\rm f}_{s}$ in each distance bin. The drop in the frequency of CN-strong stars seen in Figure 5 at $20$ kpc is also seen here.}
\end{figure}

\begin{table}
\caption{Bandstrength index definitions}
\centering
\begin{tabular}{l c c c}
\hline \hline
Index & Blue sideband & Science band & Red sideband\\
\hline
$S(3839)$ & - & $3846$-$3883\hbox{\AA}$ & $3883$-$3916\hbox{\AA}$\\
$S(CH)$ & $4050$-$4100\hbox{\AA}$ & $4280$-$4320\hbox{\AA}$ & $4330$-$4350\hbox{\AA}$\\
\hline
\end{tabular}
\end{table}

To further investigate this result, we visualize the relationship
between the frequency of CN-strong stars and Galactocentric distance
in two ways: Figure 7 shows ${\rm f}_{s}$, the ratio of CN-strong to
CN-normal stars, versus Galactocentric distance for the present data
set combined with MG10. Vertical error bars represent Poisson errors in
${\rm f}_{s}$. The fraction of CN-strong stars is lower at larger ${\rm
r}_{GC}$, with a transition at $\simeq 20$ kpc: relatively small-${\rm
r}_{GC}$ stars are consistent with a value for ${\rm f}_{s}$ around
$0.04$ and relatively large-${\rm r}_{GC}$ stars are consistent with a
value for ${\rm f}_{s}$ around $0.01$. Figure 8 shows cumulative
distribution functions of $\delta S(3839)$ for stars with ${\rm r_{GC}}
< 20$ kpc (solid line) and ${\rm r_{GC}} \geq 20$ kpc (dotted line). The
cumulative distribution curve for stars with smaller Galactocentric
distances climbs continuously for $\delta S(3839) > 0.05$, while the
curve for more distant stars is flatter in that range, indicating that
there are relatively more CN-strong stars at smaller Galactocentric
distances. A one-sided Kolmogorov-Smirnov test returns a probability of
$0.15$ that the $\delta S(3839)$ distributions in these two distance
ranges were drawn from the same parent population. This result is
only a very marginal rejection of the single population hypothesis;
however, it clearly points to a reasonable possibility that two parent
populations with differing CN-strong fractions exist.

\begin{table*}
\caption{Measured and derived parameters for all program stars from
  this study and from MG10.}
\centering
\begin{tabular}{c c c c c c c c c c}
\hline \hline
Plate & MJD & FiberID & $S(3839)$ & $S(CH)$ & CN class & ${\rm M}_{r}$
& D (kpc) & R$_{GC}$ (kpc) & Survey\\
\hline
3289 & 54910 & 132 & -0.026 & 0.716 & Normal & 0.31 & 10.3 & 17.7 & SEGUE-2\\
3373 & 54940 & 51 & 0.004 & 0.760 & Normal & -0.12 & 12.9 & 17.2 & SEGUE-2\\
3308 & 54919 & 529 & -0.008 & 0.783 & Normal & 0.06 & 18.6 & 25.2 & SEGUE-2\\
3478 & 55008 & 229 & 0.014 & 0.754 & Normal & -0.23 & 18.0 & 25.5 & SEGUE-2\\
3178 & 54848 & 269 & 0.035 & 0.763 & Normal & 0.22 & 18.2 & 22.6 & SEGUE-2\\
3128 & 54776 & 104 & 0.025 & 0.710 & Normal & 0.01 & 25.3 & 30.3 & SEGUE-2\\
3178 & 54848 & 55 & -0.153 & 0.789 & Normal & 0.80 & 15.9 & 20.5 & SEGUE-2\\
3181 & 54860 & 73 & -0.106 & 0.819 & Normal & -0.32 & 33.8 & 38.2 & SEGUE-2\\
3214 & 54866 & 14 & -0.083 & 0.798 & Normal & 0.40 & 14.1 & 19.3 & SEGUE-2\\
3298 & 54924 & 48 & -0.105 & 0.784 & Normal & 0.33 & 20.4 & 27.6 & SEGUE-2\\
\hline
\end{tabular}
\end{table*}

We interpret this drop-off in the frequency of CN-strong stars with Galactocentric distance as a
possible sign of a transition from the inner-halo population to the
outer-halo population (as discussed by, e.g., Carollo et al. 2007;
2010\nocite{CB07}\nocite{CB10}). Note that this also corresponds to
the transition zone reported by \citet{dJY10}, based on fits to Hess
diagrams from the SEGUE vertical photometry stripes. Recent
theoretical studies of galaxy formation (e.g., Oser et
al. 2010\nocite{OO10}; Font et al. 2011\nocite{FM11}) predict that
stars formed {\it in situ} are the dominant population at small
Galactocentric radii, while stars accreted during minor mergers are
the majority at larger Galactocentric distances. Studies of abundances
in individual stars in dwarf galaxies do not often report stars with
light-element abundances similar to second-generation globular cluster
stars. In \citet{SVT03} there is only one star out of the 15 surveyed
(Fornax 21) with even mildly elevated [Na/Fe], \citet{SA04} found
fairly low [Na/Fe] abundances in the three stars they observed in Ursa
Minor, and \citet{GS05} find the same result for nine RGB stars in
Sculptor. \citet{KG08} also report one star in Carina (of eight with
measured sodium abundance) with moderate [Na/Fe], and \citet{LH10}
report one star in Fornax (of 70 with measured sodium abundances) with
moderate [Na/Fe], but neither study finds any stars with the level of
sodium abundance enhancement typically found in second-generation
globular cluster stars (e.g., Carretta et al. 2009\nocite{CBG09}). 

\begin{figure}
\resizebox{\hsize}{!}{\includegraphics{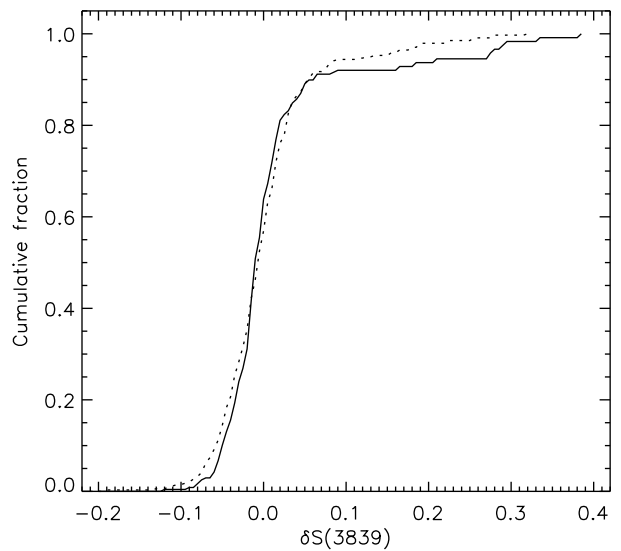}}
\caption[]{Cumulative distribution functions for $\delta S(3839)$ for
  stars with ${\rm r_{GC}} < 20$ kpc (solid line) and ${\rm
    r_{GC}}\geq 20$ kpc. The fraction of stars rises more quickly at
  large values of $\delta S(3839)$ for stars with smaller
  Galactocentric radii, indicating that CN-strong stars are more
  common in the inner halo.}
\end{figure}

A schematic picture of the
Galaxy in which the inner halo has significant contributions from
globular clusters, while the outer halo is dominated by stars that
formed in dwarf galaxies, fits reasonably well with our observation that
a higher fraction of stars in the inner halo have light-element
abundances similar to those of second-generation globular cluster stars.
Spectroscopic surveys currently underway (such as APOGEE; Allende Prieto
et al. 2008\nocite{APM08}), surveys presently being planned as
companions to the Gaia mission (e.g., Cacciari et al. 2009\nocite{C09}),
and studies based on nearby stars with accurate proper motions and halo
kinematics (such as, e.g., Carollo et al. 2010\nocite{CB10}) will
be important ways to explore this question of inner/outer halo
origins further: all of them expand the effective search space,
either by surveying to larger distances or by focusing on stars with
large apogalactic distances. 

\section{Interpretation}

The homogeneity of metallicity in present-day globular clusters seems to
favor massive models of globular cluster formation, while the observed
range in oxygen and sodium abundances (see, e.g., Carretta et al.
2009\nocite{CBG09}) indicates that dilution of feedback material and
incomplete mixing of the gas that makes up the second stellar generation
play a central role in setting final light-element abundances for
individual stars. In order to evaluate the veracity of various
globular cluster formation models, it is necessary to explore their
implications for the formation of the Galactic stellar halo. Massive
models, which require a large fraction of first-generation stars to be
lost to the halo field at early times, will produce a large enhancement
between $f_{h}^{2G}$ and $f_{h}^{GC}$. 

The recent paper of \citet{SC11} considers this question from a
theoretical standpoint, using the globular cluster self-enrichment model
of \citet{DM07} to calculate the mass in first-generation stars needed
to produce a significant second generation. In the \citet{DM07} model,
rapidly rotating massive stars provide chemical feedback between the
first and second generations, and \citet{SC11} explore the effects of
varying the slope of the high-mass end of the stellar initial mass
function (IMF) on the resulting second-generation population. Requiring
that the present-day ratio of first- to second-generation stars be 1:2,
the authors calculate that initial globular cluster masses must have
been 8 to 10 times their present-day values for a Salpeter IMF. They
then compare the mass in first-generation stars that was lost from
globular clusters at early times to the mass of the present-day Galactic
stellar halo. Since roughly $35\%$ of the mass in a Salpeter IMF is in
low-mass, long-lived stars, and the present-day Galactic globular
cluster system comprises $2\%$ of the mass of the halo, they estimate
that escaped first-generation globular cluster stars make up $5-8\%$ of
the mass of the halo.

\citet{SC11} then consider the origin of the second-generation stars found
in the halo by MG10 and by \citet{CBG10b}, specifically whether they
can reasonably be understood as having been lost to the halo
during the strong early mass loss that removed roughly $90\%$ of the
first-generation stars or whether they were transferred during later
episodes of globular cluster tidal disruption. Folding in some
assumptions involving the initial cluster mass function (ICMF), they conclude
that all second-generation stars in the halo were lost at early
times. This raises the fraction of cluster mass that was lost in this
early phase, which necessarily increases the initial cluster mass to
15 to 25 times the present-day mass and also raises the fraction of
halo stars that originally formed in globular clusters to as much as
$20\%$. 

However, given the observed examples of globular cluster dissolution
in progress, e.g., Palomar 5 (Rockosi et al. 2002\nocite{ROG02};
Odenkirchen et al. 2003\nocite{OGD03}), NGC 5466 \citep{BEI06} and the
GD-1 stellar stream \citep{GD06}, as well as the relative emptiness of
the more inhospitable regions of the \citet{GO97} ``vital diagram'', we
expect cluster dissolution to be the dominant source of halo field stars
with second-generation abundance patterns. Theoretical studies of the
ICMF, early star cluster mass loss, and the evolution of the cluster
mass function (e.g., Baumgardt et al. 2008\nocite{BK08}; Parmentier et
al. 2009\nocite{PG09}; Decressin et al. 2010\nocite{DB10}) all suggest
that far more star clusters were initially formed than have survived to
the present day, especially at the lower end of the ICMF. The study of
\citet{JG10}, which used matched-filter techniques to search for faint
tidal tails around Galactic globular clusters in the SDSS imaging
footprint, found stars with cluster-like photometry outside the tidal
radii of a number of clusters without pronounced tidal tails, indicating
that stars continue to escape from globular clusters even in the absence
of dramatic tidal features. In the limit where cluster dissolution is
the sole source of second-generation field stars, we can estimate the
number of present-day globular clusters that would need to be disrupted
to provide the observed $2.5\%$ of halo stars with second-generation
chemistry, ${\rm N_{d}}$. For a typical present-day 1:1 ratio of first-
to second-generation stars, and assuming a stellar halo mass of
$10^{9}{\rm M_{\odot}}$ (e.g., Freeman \& Bland-Hawthorn
2002\nocite{FB02}) and a typical present-day globular cluster mass of $5
\times 10^{5} {\rm M_{\odot}}$, ${\rm N_{d}} \simeq \frac{2 \times {\rm
f_{h}^{2G}} \times {\rm M_{halo}}}{\langle {\rm M_{gc}} \rangle} \simeq
100$. Since the current number of of globular clusters is on the order
of $150$, this means that the initial population of globular clusters
was significantly larger, but this is not an outrageous claim:
\citet{Mv05} calculate that the ratio between the initial and
present-day number of globular clusters in the Galaxy is $3/2$, whereas
we derive $5/3$. 

Whatever the details, it seems clear that first-generation stars lost
from globular clusters at early times are an important contribution to
the construction of the Galactic halo, if two-generation self-enrichment
scenarios for globular cluster formation are indeed the correct model.
The models of \citet{DM07}, \citet{DVD08}, \citet{CBG10b}, and
\citet{VM10} all predict that $90\%$ of the first generation ought to
be lost from globular clusters at early times, while \citet{C11}
estimates that as much as $95\%$ of the first generation ought to be
lost. The mass of first-generation stars lost during this phase by the
${\rm N_{GC}}$ globular clusters in the present-day system can be
expressed as ${\rm M_{1G}^{lost}}={\rm M_{GC~system}}\times \frac{1}{2}
\times (\frac{1}{1-{\rm f_{lost}}}-1)$. Considering the early
contributions of first-generation stars from globular clusters that have
survived to the present day as well as the clusters that have completely
dissolved, we estimate that ${\rm f_{h}^{GC}}$ has a minimum value of
$\frac{\rm M_{1G}^{lost}}{\rm M_{halo}} + (\frac{\rm N_{d}}{\rm N_{GC}}
\times \frac{\rm M_{1G}^{lost}}{\rm M_{halo}}) + {\rm M_{GC~system}} =
2\% \times (\frac{9}{2} + \frac{12}{3}) = 17\%$ of halo field stars,
with both first- and second-generation abundance patterns, originally
formed within globular clusters, plus an additional unknown contribution
of first-generation stars from clusters too low-mass to self-enrich that
have dispersed by the present day.

\section{Future challenges}

Based on the current two-generation models for globular cluster
formation, the early phase of globular cluster formation must have
contributed a significant number of first-generation stars to the
Galactic halo. From the data presented herein, by MG10, and by
\citet{CBG10b}, it is also clear that second-generation globular
cluster stars are a component of the halo field. Further
investigations of the process of globular cluster formation will help
to clarify the role that globular clusters played in the early assembly
of the Galactic halo. There are two specific directions that new
observations could take that would be particularly insightful:
investigating the relationship between the minimum mass for cluster
self-enrichment and the larger galactic environment, and searching for
star clusters still in the process of self-enrichment.

These studies will necessarily require observations of star clusters
outside the Milky Way, as a way to compare cluster formation in
different large-scale environments. Observing very young star clusters
may offer the opportunity to see the process of self-enrichment in
progress. \citet{CBG10b} suggest that clusters in the process of violent
relaxation (at a few $\times 10^{7}$ years old) should have a central
compact cluster, an extended, non-bound halo and outflowing gas. It is
suggested in \citet{C11} that the extended clusters discovered in M31 by
\citet{HT05} are in this state (see also Huxor et al. 2011\nocite{HF11}). 
If this is the case, the stars at large radius in the extended clusters
should have almost entirely first-generation chemistry, and strong
radial gradients in light-element abundances, most readily observable in
CN and CH molecular features, should be present.

To explore the question of the minimum cluster mass for self-enrichment,
it would be useful to consider star clusters in lower-mass galaxies.
\citet{CS11} claim that the weaker tidal field of the Large Magellanic
Cloud (LMC) ought to permit the formation of a second stellar generation
at lower cluster mass than in the Milky Way. Spectroscopic observations
of red giant-branch stars in intermediate-age populous clusters in the
LMC would be particularly interesting because recent photometric work
(e.g., Mackey et al. 2008\nocite{MBN08}; Milone et al.
2009\nocite{MBP09}; Goudfrooij et al. 2011\nocite{GP11}) has uncovered
broadened or split main-sequence turnoffs in a large fraction of them.
Since they are much younger than Galactic globular clusters (typically
$1$-$4$ Gyr), age differences on the order of a few hundred million
years are visible at the turnoff, and this age difference is
suggestively similar to the age difference expected between first- and
second-generation stars in Galactic globular clusters. Obtaining spectra
of turnoff stars at the distance of the LMC would be extremely
difficult -- indeed, high-resolution spectroscopic studies of RGB stars in
these clusters are limited to fairly small samples (e.g., Mucciarelli et
al. 2008\nocite{MCO08}). However, a lower-resolution study using a
multiobject spectrograph could collect CN and CH data analogous to the
data presented in this paper, and would allow a search for star-to-star
variations in carbon and nitrogen abundance for a larger data set per
cluster. A modified version of the SSPP has been developed for
application to spectra with resolving power as low as $R \sim 1000$),
and is already being used with a variety of low-resolution data from
non-SDSS sources (e.g., see Li et al. 2010\nocite{LC10}; Humphreys et
al 2011\nocite{HB11}). These data would also permit an investigation
of theoretical claims (e.g., Conroy \& Spergel 2011\nocite{CS11}) that
lower-mass galaxies permit the formation of multiple stellar generations in
lower-mass star clusters, by comparing the CN-CH behavior of
intermediate-age LMC clusters across a range of masses.

\newpage
\bibliography{cnsdr8}

\begin{thebibliography}{85}
\expandafter\ifx\csname natexlab\endcsname\relax\def\natexlab#1{#1}\fi

\bibitem[{{Abazajian} {et~al.}(2004){Abazajian}, {Adelman-McCarthy},
  {Ag{\"u}eros}, {Allam}, {Anderson}, {Anderson}, {Annis}, {Bahcall}, {Baldry},
  \& {Bastian}}]{DR2}
{Abazajian}, K., {Adelman-McCarthy}, J.~K., {Ag{\"u}eros}, M.~A., {et~al.}
  2004, \aj, 128, 502

\bibitem[{{Abazajian} {et~al.}(2005){Abazajian}, {Adelman-McCarthy},
  {Ag{\"u}eros}, {Allam}, {Anderson}, {Anderson}, {Annis}, {Bahcall}, {Baldry},
  \& {Bastian}}]{DR3}
{Abazajian}, K., {Adelman-McCarthy}, J.~K., {Ag{\"u}eros}, M.~A., {et~al.}
  2005, \aj, 129, 1755

\bibitem[{{Abazajian} {et~al.}(2003){Abazajian}, {Adelman-McCarthy},
  {Ag{\"u}eros}, {Allam}, {Anderson}, {Annis}, {Bahcall}, {Baldry}, {Bastian},
  \& {Berlind}}]{DR1}
{Abazajian}, K., {Adelman-McCarthy}, J.~K., {Ag{\"u}eros}, M.~A., {et~al.}
  2003, \aj, 126, 2081

\bibitem[{{Abazajian} {et~al.}(2009){Abazajian}, {Adelman-McCarthy},
  {Ag{\"u}eros}, {Allam}, {Allende Prieto}, {An}, {Anderson}, {Anderson},
  {Annis}, \& {Bahcall}}]{DR7}
{Abazajian}, K.~N., {Adelman-McCarthy}, J.~K., {Ag{\"u}eros}, M.~A., {et~al.}
  2009, \apjs, 182, 543

\bibitem[{{Adelman-McCarthy} {et~al.}(2008){Adelman-McCarthy}, {Ag{\"u}eros},
  {Allam}, {Allende Prieto}, {Anderson}, {Anderson}, {Annis}, {Bahcall},
  {Bailer-Jones}, \& {Baldry}}]{DR6}
{Adelman-McCarthy}, J.~K., {Ag{\"u}eros}, M.~A., {Allam}, S.~S., {et~al.} 2008,
  \apjs, 175, 297

\bibitem[{{Adelman-McCarthy} {et~al.}(2007){Adelman-McCarthy}, {Ag{\"u}eros},
  {Allam}, {Anderson}, {Anderson}, {Annis}, {Bahcall}, {Bailer-Jones},
  {Baldry}, \& {Barentine}}]{DR5}
{Adelman-McCarthy}, J.~K., {Ag{\"u}eros}, M.~A., {Allam}, S.~S., {et~al.} 2007,
  \apjs, 172, 634

\bibitem[{{Adelman-McCarthy} {et~al.}(2006){Adelman-McCarthy}, {Ag{\"u}eros},
  {Allam}, {Anderson}, {Anderson}, {Annis}, {Bahcall}, {Baldry}, {Barentine},
  \& {Berlind}}]{DR4}
{Adelman-McCarthy}, J.~K., {Ag{\"u}eros}, M.~A., {Allam}, S.~S., {et~al.} 2006,
  \apjs, 162, 38

\bibitem[{{Aihara} {et~al.}(2011){Aihara}, {Allende Prieto}, {An}, {Anderson},
  {Aubourg}, {Balbinot}, {Beers}, {Berlind}, {Bickerton}, \& {Bizyaev}}]{DR8}
{Aihara}, H., {Allende Prieto}, C., {An}, D., {et~al.} 2011, \apjs, 193, 29

\bibitem[{{Allende Prieto} {et~al.}(2008{\natexlab{a}}){Allende Prieto},
  {Majewski}, {Schiavon}, {Cunha}, {Frinchaboy}, {Holtzman}, {Johnston},
  {Shetrone}, {Skrutskie}, {Smith}, \& {Wilson}}]{APM08}
{Allende Prieto}, C., {Majewski}, S.~R., {Schiavon}, R., {et~al.}
  2008{\natexlab{a}}, Astronomische Nachrichten, 329, 1018

\bibitem[{{Allende Prieto} {et~al.}(2008{\natexlab{b}}){Allende Prieto},
  {Sivarani}, {Beers}, {Lee}, {Koesterke}, {Shetrone}, {Sneden}, {Lambert},
  {Wilhelm}, {Rockosi}, {Lai}, {Yanny}, {Ivans}, {Johnson}, {Aoki},
  {Bailer-Jones}, \& {Re Fiorentin}}]{AP08}
{Allende Prieto}, C., {Sivarani}, T., {Beers}, T.~C., {et~al.}
  2008{\natexlab{b}}, \aj, 136, 2070

\bibitem[{{Baumgardt} {et~al.}(2008){Baumgardt}, {Kroupa}, \&
  {Parmentier}}]{BK08}
{Baumgardt}, H., {Kroupa}, P., \& {Parmentier}, G. 2008, \mnras, 384, 1231

\bibitem[{{Bekki} {et~al.}(2007){Bekki}, {Campbell}, {Lattanzio}, \&
  {Norris}}]{BC07}
{Bekki}, K., {Campbell}, S.~W., {Lattanzio}, J.~C., \& {Norris}, J.~E. 2007,
  \mnras, 377, 335

\bibitem[{{Bell} \& {Dickens}(1980)}]{BD80}
{Bell}, R.~A. \& {Dickens}, R.~J. 1980, \apj, 242, 657

\bibitem[{{Belokurov} {et~al.}(2006){Belokurov}, {Evans}, {Irwin}, {Hewett}, \&
  {Wilkinson}}]{BEI06}
{Belokurov}, V., {Evans}, N.~W., {Irwin}, M.~J., {Hewett}, P.~C., \&
  {Wilkinson}, M.~I. 2006, \apjl, 637, L29

\bibitem[{{Briley} {et~al.}(2004){Briley}, {Cohen}, \& {Stetson}}]{BCS04}
{Briley}, M.~M., {Cohen}, J.~G., \& {Stetson}, P.~B. 2004, \aj, 127, 1579

\bibitem[{{Cacciari}(2009)}]{C09}
{Cacciari}, C. 2009, \memsai, 80, 97

\bibitem[{{Carollo} {et~al.}(2010){Carollo}, {Beers}, {Chiba}, {Norris},
  {Freeman}, {Lee}, {Ivezi{\'c}}, {Rockosi}, \& {Yanny}}]{CB10}
{Carollo}, D., {Beers}, T.~C., {Chiba}, M., {et~al.} 2010, \apj, 712, 692

\bibitem[{{Carollo} {et~al.}(2007){Carollo}, {Beers}, {Lee}, {Chiba}, {Norris},
  {Wilhelm}, {Sivarani}, {Marsteller}, {Munn}, {Bailer-Jones}, {Fiorentin}, \&
  {York}}]{CB07}
{Carollo}, D., {Beers}, T.~C., {Lee}, Y.~S., {et~al.} 2007, \nat, 450, 1020

\bibitem[{{Carretta} {et~al.}(2009){Carretta}, {Bragaglia}, {Gratton},
  {Lucatello}, {Catanzaro}, {Leone}, {Bellazzini}, {Claudi}, {D'Orazi},
  {Momany}, {Ortolani}, {Pancino}, {Piotto}, {Recio-Blanco}, \&
  {Sabbi}}]{CBG09}
{Carretta}, E., {Bragaglia}, A., {Gratton}, R.~G., {et~al.} 2009, \aap, 505,
  117

\bibitem[{{Carretta} {et~al.}(2010){Carretta}, {Bragaglia}, {Gratton},
  {Recio-Blanco}, {Lucatello}, {D'Orazi}, \& {Cassisi}}]{CBG10b}
{Carretta}, E., {Bragaglia}, A., {Gratton}, R.~G., {et~al.} 2010, \aap, 516,
  A55

\bibitem[{{Conroy}(2011)}]{C11}
{Conroy}, C. 2011, ArXiv e-prints: 1101.2208

\bibitem[{{Conroy} \& {Spergel}(2011)}]{CS11}
{Conroy}, C. \& {Spergel}, D.~N. 2011, \apj, 726, 36

\bibitem[{{de Jong} {et~al.}(2010){de Jong}, {Yanny}, {Rix}, {Dolphin},
  {Martin}, \& {Beers}}]{dJY10}
{de Jong}, J.~T.~A., {Yanny}, B., {Rix}, H.-W., {et~al.} 2010, \apj, 714, 663

\bibitem[{{de Mink} {et~al.}(2009){de Mink}, {Pols}, {Langer}, \&
  {Izzard}}]{DMP09}
{de Mink}, S.~E., {Pols}, O.~R., {Langer}, N., \& {Izzard}, R.~G. 2009, \aap,
  507, L1

\bibitem[{{de Silva} {et~al.}(2009){de Silva}, {Gibson}, {Lattanzio}, \&
  {Asplund}}]{dSG09}
{de Silva}, G.~M., {Gibson}, B.~K., {Lattanzio}, J., \& {Asplund}, M. 2009,
  \aap, 500, L25

\bibitem[{{Decressin} {et~al.}(2010){Decressin}, {Baumgardt}, {Charbonnel}, \&
  {Kroupa}}]{DB10}
{Decressin}, T., {Baumgardt}, H., {Charbonnel}, C., \& {Kroupa}, P. 2010, \aap,
  516, A73

\bibitem[{{Decressin} {et~al.}(2007){Decressin}, {Meynet}, {Charbonnel},
  {Prantzos}, \& {Ekstr{\"o}m}}]{DM07}
{Decressin}, T., {Meynet}, G., {Charbonnel}, C., {Prantzos}, N., \&
  {Ekstr{\"o}m}, S. 2007, \aap, 464, 1029

\bibitem[{{D'Ercole} {et~al.}(2011){D'Ercole}, {D'Antona}, \&
  {Vesperini}}]{DDV11}
{D'Ercole}, A., {D'Antona}, F., \& {Vesperini}, E. 2011, \mnras, 415, 1304

\bibitem[{{D'Ercole} {et~al.}(2008){D'Ercole}, {Vesperini}, {D'Antona},
  {McMillan}, \& {Recchi}}]{DVD08}
{D'Ercole}, A., {Vesperini}, E., {D'Antona}, F., {McMillan}, S.~L.~W., \&
  {Recchi}, S. 2008, \mnras, 391, 825

\bibitem[{{Dotter} {et~al.}(2008){Dotter}, {Chaboyer}, {Jevremovi{\'c}},
  {Kostov}, {Baron}, \& {Ferguson}}]{DC08}
{Dotter}, A., {Chaboyer}, B., {Jevremovi{\'c}}, D., {et~al.} 2008, \apjs, 178,
  89

\bibitem[{{Eisenstein} {et~al.}(2011){Eisenstein}, {Weinberg}, {Agol},
  {Aihara}, {Allende Prieto}, {Anderson}, {Arns}, {Aubourg}, {Bailey},
  {Balbinot}, \& et~al.}]{SDSSIII}
{Eisenstein}, D.~J., {Weinberg}, D.~H., {Agol}, E., {et~al.} 2011, \aj, 142, 72

\bibitem[{{Font} {et~al.}(2011){Font}, {McCarthy}, {Crain}, {Theuns}, {Schaye},
  {Wiersma}, \& {Dalla Vecchia}}]{FM11}
{Font}, A.~S., {McCarthy}, I.~G., {Crain}, R.~A., {et~al.} 2011, \mnras, 1162

\bibitem[{{Freeman} \& {Bland-Hawthorn}(2002)}]{FB02}
{Freeman}, K. \& {Bland-Hawthorn}, J. 2002, \araa, 40, 487

\bibitem[{{Fukugita} {et~al.}(1996){Fukugita}, {Ichikawa}, {Gunn}, {Doi},
  {Shimasaku}, \& {Schneider}}]{FI96}
{Fukugita}, M., {Ichikawa}, T., {Gunn}, J.~E., {et~al.} 1996, \aj, 111, 1748

\bibitem[{{Geisler} {et~al.}(2005){Geisler}, {Smith}, {Wallerstein},
  {Gonzalez}, \& {Charbonnel}}]{GS05}
{Geisler}, D., {Smith}, V.~V., {Wallerstein}, G., {Gonzalez}, G., \&
  {Charbonnel}, C. 2005, \aj, 129, 1428

\bibitem[{{Gnedin} \& {Ostriker}(1997)}]{GO97}
{Gnedin}, O.~Y. \& {Ostriker}, J.~P. 1997, \apj, 474, 223

\bibitem[{{Goudfrooij} {et~al.}(2011){Goudfrooij}, {Puzia}, {Chandar}, \&
  {Kozhurina-Platais}}]{GP11}
{Goudfrooij}, P., {Puzia}, T.~H., {Chandar}, R., \& {Kozhurina-Platais}, V.
  2011, \apj, 737, 4

\bibitem[{{Grillmair} \& {Dionatos}(2006)}]{GD06}
{Grillmair}, C.~J. \& {Dionatos}, O. 2006, \apjl, 643, L17

\bibitem[{{Gunn} {et~al.}(1998){Gunn}, {Carr}, {Rockosi}, {Sekiguchi}, {Berry},
  {Elms}, {de Haas}, {Ivezi{\'c}}, {Knapp}, \& {Lupton}}]{GC98}
{Gunn}, J.~E., {Carr}, M., {Rockosi}, C., {et~al.} 1998, \aj, 116, 3040

\bibitem[{{Gunn} {et~al.}(2006){Gunn}, {Siegmund}, {Mannery}, {Owen}, {Hull},
  {Leger}, {Carey}, {Knapp}, {York}, \& {Boroski}}]{GS06}
{Gunn}, J.~E., {Siegmund}, W.~A., {Mannery}, E.~J., {et~al.} 2006, \aj, 131,
  2332

\bibitem[{{Harbeck} {et~al.}(2003){Harbeck}, {Smith}, \& {Grebel}}]{HSG03}
{Harbeck}, D., {Smith}, G.~H., \& {Grebel}, E.~K. 2003, \aj, 125, 197

\bibitem[{{Humphreys} {et~al.}(2011){Humphreys}, {Beers}, {Cabanela},
  {Grammer}, {Davidson}, {Lee}, \& {Larsen}}]{HB11}
{Humphreys}, R.~M., {Beers}, T.~C., {Cabanela}, J.~E., {et~al.} 2011, \aj, 141,
  131

\bibitem[{{Huxor} {et~al.}(2011){Huxor}, {Ferguson}, {Tanvir}, {Irwin},
  {Mackey}, {Ibata}, {Bridges}, {Chapman}, \& {Lewis}}]{HF11}
{Huxor}, A.~P., {Ferguson}, A.~M.~N., {Tanvir}, N.~R., {et~al.} 2011, \mnras,
  414, 770

\bibitem[{{Huxor} {et~al.}(2005){Huxor}, {Tanvir}, {Irwin}, {Ibata}, {Collett},
  {Ferguson}, {Bridges}, \& {Lewis}}]{HT05}
{Huxor}, A.~P., {Tanvir}, N.~R., {Irwin}, M.~J., {et~al.} 2005, \mnras, 360,
  1007

\bibitem[{{Jacobson} {et~al.}(2008){Jacobson}, {Friel}, \&
  {Pilachowski}}]{JFP08}
{Jacobson}, H.~R., {Friel}, E.~D., \& {Pilachowski}, C.~A. 2008, \aj, 135, 2341

\bibitem[{{Jordi} \& {Grebel}(2010)}]{JG10}
{Jordi}, K. \& {Grebel}, E.~K. 2010, \aap, 522, A71

\bibitem[{{Karakas}(2010)}]{K10}
{Karakas}, A.~I. 2010, \mnras, 403, 1413

\bibitem[{{Kayser} {et~al.}(2008){Kayser}, {Hilker}, {Grebel}, \&
  {Willemsen}}]{KHG08}
{Kayser}, A., {Hilker}, M., {Grebel}, E.~K., \& {Willemsen}, P.~G. 2008, \aap,
  486, 437

\bibitem[{{Koch} {et~al.}(2008){Koch}, {Grebel}, {Gilmore}, {Wyse}, {Kleyna},
  {Harbeck}, {Wilkinson}, \& {Wyn Evans}}]{KG08}
{Koch}, A., {Grebel}, E.~K., {Gilmore}, G.~F., {et~al.} 2008, \aj, 135, 1580

\bibitem[{{Kraft}(1994)}]{K94}
{Kraft}, R.~P. 1994, \pasp, 106, 553

\bibitem[{{Lee} {et~al.}(2008{\natexlab{a}}){Lee}, {Beers}, {Sivarani},
  {Allende Prieto}, {Koesterke}, {Wilhelm}, {Re Fiorentin}, {Bailer-Jones},
  {Norris}, {Rockosi}, {Yanny}, {Newberg}, {Covey}, {Zhang}, \& {Luo}}]{LBS08a}
{Lee}, Y.~S., {Beers}, T.~C., {Sivarani}, T., {et~al.} 2008{\natexlab{a}}, \aj,
  136, 2022

\bibitem[{{Lee} {et~al.}(2008{\natexlab{b}}){Lee}, {Beers}, {Sivarani},
  {Johnson}, {An}, {Wilhelm}, {Allende Prieto}, {Koesterke}, {Re Fiorentin},
  {Bailer-Jones}, {Norris}, {Yanny}, {Rockosi}, {Newberg}, {Cudworth}, \&
  {Pan}}]{LBS08b}
{Lee}, Y.~S., {Beers}, T.~C., {Sivarani}, T., {et~al.} 2008{\natexlab{b}}, \aj,
  136, 2050

\bibitem[{{Letarte} {et~al.}(2010){Letarte}, {Hill}, {Tolstoy}, {Jablonka},
  {Shetrone}, {Venn}, {Spite}, {Irwin}, {Battaglia}, {Helmi}, {Primas},
  {Fran{\c c}ois}, {Kaufer}, {Szeifert}, {Arimoto}, \& {Sadakane}}]{LH10}
{Letarte}, B., {Hill}, V., {Tolstoy}, E., {et~al.} 2010, \aap, 523, A17

\bibitem[{{Li} {et~al.}(2010){Li}, {Christlieb}, {Sch{\"o}rck}, {Norris},
  {Bessell}, {Yong}, {Beers}, {Lee}, {Frebel}, \& {Zhao}}]{LC10}
{Li}, H.~N., {Christlieb}, N., {Sch{\"o}rck}, T., {et~al.} 2010, \aap, 521, A10

\bibitem[{{Mackey} {et~al.}(2008){Mackey}, {Broby Nielsen}, {Ferguson}, \&
  {Richardson}}]{MBN08}
{Mackey}, A.~D., {Broby Nielsen}, P., {Ferguson}, A.~M.~N., \& {Richardson},
  J.~C. 2008, \apjl, 681, L17

\bibitem[{{Mackey} \& {van den Bergh}(2005)}]{Mv05}
{Mackey}, A.~D. \& {van den Bergh}, S. 2005, \mnras, 360, 631

\bibitem[{{Marks} \& {Kroupa}(2010)}]{MK10}
{Marks}, M. \& {Kroupa}, P. 2010, \mnras, 406, 2000

\bibitem[{{Martell}(2011)}]{M11}
{Martell}, S.~L. 2011, Astronomische Nachrichten, 332, 467

\bibitem[{{Martell} \& {Grebel}(2010)}]{MG10}
{Martell}, S.~L. \& {Grebel}, E.~K. 2010, \aap, 519, A14

\bibitem[{{Martell} \& {Smith}(2009)}]{MS09}
{Martell}, S.~L. \& {Smith}, G.~H. 2009, \pasp, 121, 577

\bibitem[{{Martell} {et~al.}(2008){Martell}, {Smith}, \& {Briley}}]{MSB08}
{Martell}, S.~L., {Smith}, G.~H., \& {Briley}, M.~M. 2008, \pasp, 120, 7

\bibitem[{{Milone} {et~al.}(2009){Milone}, {Bedin}, {Piotto}, \&
  {Anderson}}]{MBP09}
{Milone}, A.~P., {Bedin}, L.~R., {Piotto}, G., \& {Anderson}, J. 2009, \aap,
  497, 755

\bibitem[{{Mucciarelli} {et~al.}(2008){Mucciarelli}, {Carretta}, {Origlia}, \&
  {Ferraro}}]{MCO08}
{Mucciarelli}, A., {Carretta}, E., {Origlia}, L., \& {Ferraro}, F.~R. 2008,
  \aj, 136, 375

\bibitem[{{Mucciarelli} {et~al.}(2011){Mucciarelli}, {Cristallo}, {Brocato},
  {Pasquini}, {Straniero}, {Caffau}, {Raimondo}, {Kaufer}, {Musella}, {Ripepi},
  {Romaniello}, \& {Walker}}]{MCB11}
{Mucciarelli}, A., {Cristallo}, S., {Brocato}, E., {et~al.} 2011, \mnras, 413,
  837

\bibitem[{{Norris} \& {Cottrell}(1979)}]{NC79}
{Norris}, J. \& {Cottrell}, P.~L. 1979, \apjl, 229, L69

\bibitem[{{Norris} {et~al.}(1981){Norris}, {Cottrell}, {Freeman}, \& {Da
  Costa}}]{N81}
{Norris}, J., {Cottrell}, P.~L., {Freeman}, K.~C., \& {Da Costa}, G.~S. 1981,
  \apj, 244, 205

\bibitem[{{Norris} {et~al.}(1984){Norris}, {Freeman}, \& {Da Costa}}]{N84}
{Norris}, J., {Freeman}, K.~C., \& {Da Costa}, G.~S. 1984, \apj, 277, 615

\bibitem[{{Odenkirchen} {et~al.}(2003){Odenkirchen}, {Grebel}, {Dehnen}, {Rix},
  {Yanny}, {Newberg}, {Rockosi}, {Mart{\'{\i}}nez-Delgado}, {Brinkmann}, \&
  {Pier}}]{OGD03}
{Odenkirchen}, M., {Grebel}, E.~K., {Dehnen}, W., {et~al.} 2003, \aj, 126, 2385

\bibitem[{{Oser} {et~al.}(2010){Oser}, {Ostriker}, {Naab}, {Johansson}, \&
  {Burkert}}]{OO10}
{Oser}, L., {Ostriker}, J.~P., {Naab}, T., {Johansson}, P.~H., \& {Burkert}, A.
  2010, \apj, 725, 2312

\bibitem[{{Pancino} {et~al.}(2010){Pancino}, {Rejkuba}, {Zoccali}, \&
  {Carrera}}]{PR10}
{Pancino}, E., {Rejkuba}, M., {Zoccali}, M., \& {Carrera}, R. 2010, \aap, 524,
  A44

\bibitem[{{Parmentier} {et~al.}(2009){Parmentier}, {Goodwin}, {Kroupa}, \&
  {Baumgardt}}]{PG09}
{Parmentier}, G., {Goodwin}, S.~P., {Kroupa}, P., \& {Baumgardt}, H. 2009,
  \apss, 324, 327

\bibitem[{{Parmentier} {et~al.}(1999){Parmentier}, {Jehin}, {Magain},
  {Neuforge}, {Noels}, \& {Thoul}}]{PJM99}
{Parmentier}, G., {Jehin}, E., {Magain}, P., {et~al.} 1999, \aap, 352, 138

\bibitem[{{Pier} {et~al.}(2003){Pier}, {Munn}, {Hindsley}, {Hennessy}, {Kent},
  {Lupton}, \& {Ivezi{\'c}}}]{PM03}
{Pier}, J.~R., {Munn}, J.~A., {Hindsley}, R.~B., {et~al.} 2003, \aj, 125, 1559

\bibitem[{{Priestley} {et~al.}(2011){Priestley}, {Ruffert}, \&
  {Salaris}}]{PR11}
{Priestley}, W., {Ruffert}, M., \& {Salaris}, M. 2011, \mnras, 411, 1935

\bibitem[{{Rockosi} {et~al.}(2002){Rockosi}, {Odenkirchen}, {Grebel}, {Dehnen},
  {Cudworth}, {Gunn}, {York}, {Brinkmann}, {Hennessy}, \& {Ivezi{\'c}}}]{ROG02}
{Rockosi}, C.~M., {Odenkirchen}, M., {Grebel}, E.~K., {et~al.} 2002, \aj, 124,
  349

\bibitem[{{Sadakane} {et~al.}(2004){Sadakane}, {Arimoto}, {Ikuta}, {Aoki},
  {Jablonka}, \& {Tajitsu}}]{SA04}
{Sadakane}, K., {Arimoto}, N., {Ikuta}, C., {et~al.} 2004, \pasj, 56, 1041

\bibitem[{{Schaerer} \& {Charbonnel}(2011)}]{SC11}
{Schaerer}, D. \& {Charbonnel}, C. 2011, \mnras, 413, 2297

\bibitem[{{Shetrone} {et~al.}(2003){Shetrone}, {Venn}, {Tolstoy}, {Primas},
  {Hill}, \& {Kaufer}}]{SVT03}
{Shetrone}, M., {Venn}, K.~A., {Tolstoy}, E., {et~al.} 2003, \aj, 125, 684

\bibitem[{{Smolinski} {et~al.}(2011{\natexlab{a}}){Smolinski}, {Lee}, {Beers},
  {An}, {Bickerton}, {Johnson}, {Loomis}, {Rockosi}, {Sivarani}, \&
  {Yanny}}]{SL11}
{Smolinski}, J.~P., {Lee}, Y.~S., {Beers}, T.~C., {et~al.} 2011{\natexlab{a}},
  \aj, 141, 89

\bibitem[{{Smolinski} {et~al.}(2011{\natexlab{b}}){Smolinski}, {Martell},
  {Beers}, \& {Lee}}]{SM11}
{Smolinski}, J.~P., {Martell}, S.~L., {Beers}, T.~C., \& {Lee}, Y.~S.
  2011{\natexlab{b}}, \aj, 142, 126

\bibitem[{{Stoughton} {et~al.}(2002){Stoughton}, {Lupton}, {Bernardi},
  {Blanton}, {Burles}, {Castander}, {Connolly}, {Eisenstein}, {Frieman}, \&
  {Hennessy}}]{SL02}
{Stoughton}, C., {Lupton}, R.~H., {Bernardi}, M., {et~al.} 2002, \aj, 123, 485

\bibitem[{{Suntzeff}(1981)}]{S81}
{Suntzeff}, N.~B. 1981, \apjs, 47, 1

\bibitem[{{Vesperini} {et~al.}(2010){Vesperini}, {McMillan}, {D'Antona}, \&
  {D'Ercole}}]{VM10}
{Vesperini}, E., {McMillan}, S.~L.~W., {D'Antona}, F., \& {D'Ercole}, A. 2010,
  \apjl, 718, L112

\bibitem[{{Yanny} {et~al.}(2009){Yanny}, {Rockosi}, {Newberg}, {Knapp},
  {Adelman-McCarthy}, {Alcorn}, {Allam}, {Allende Prieto}, {An}, \&
  {Anderson}}]{YR09}
{Yanny}, B., {Rockosi}, C., {Newberg}, H.~J., {et~al.} 2009, \aj, 137, 4377

\bibitem[{{York} {et~al.}(2000){York}, {Adelman}, {Anderson}, {Anderson},
  {Annis}, {Bahcall}, {Bakken}, {Barkhouser}, {Bastian}, \& {Berman}}]{Y00}
{York}, D.~G., {Adelman}, J., {Anderson}, Jr., J.~E., {et~al.} 2000, \aj, 120,
  1579

\end{thebibliography}

\newpage

\begin{acknowledgements}
SLM was supported by Sonderforschungsbereich SFB 881 ``The Milky Way
System'' (subprojects A2 and A5) of the German
Research Foundation (DFG). JPS and TCB acknowledge partial funding of
this work from grants PHY 02-1783 and PHY 08-22648: Physics Frontiers
Center/Joint Institute for Nuclear Astrophysics (JINA), awarded by the
National Science Foundation.

Funding for SDSS-III has been provided by the Alfred P. Sloan Foundation, the Participating Institutions, the National Science Foundation, and the U.S. Department of Energy. The SDSS-III web site is http://www.sdss3.org/.

SDSS-III is managed by the Astrophysical Research Consortium for the Participating Institutions of the SDSS-III Collaboration including the University of Arizona, the Brazilian Participation Group, Brookhaven National Laboratory, University of Cambridge, University of Florida, the French Participation Group, the German Participation Group, the Instituto de Astrofisica de Canarias, the Michigan State/Notre Dame/JINA Participation Group, Johns Hopkins University, Lawrence Berkeley National Laboratory, Max Planck Institute for Astrophysics, New Mexico State University, New York University, Ohio State University, Pennsylvania State University, University of Portsmouth, Princeton University, the Spanish Participation Group, University of Tokyo, University of Utah, Vanderbilt University, University of Virginia, University of Washington, and Yale University.
\end{acknowledgements}

\end{document}